%% file: ixc.tex
\documentclass[11pt]{article}
\usepackage{jcappub}
\usepackage{graphicx}
\usepackage{natbib}
\usepackage{amsmath}
\usepackage{color,colordvi}
\usepackage{multirow}
\usepackage[mediumspace,Gray,squaren]{SIunits}
\usepackage{hyperref}

\input{aas_macros}
\definecolor{darkblue}{rgb}{0,0,0.3}
\definecolor{darkgreen}{rgb}{0,0.3,0}
\hypersetup{
  pdftitle = {IRIS x MaxBCG Clusters},
  pdfauthor = {Hincks, Hajian & Addison},
  colorlinks,%
  citecolor=darkblue,%
  filecolor=black,%
  linkcolor=blue,%
  urlcolor=blue
}

\newcommand{\bm}[1]{\textbf{#1}}
\addunit{\jansky}{\ensuremath{\mathrm{Jy}}}
\addunit{\degword}{\ensuremath{\mathrm{deg}}}

\begin{document}

\title{A High-Significance Measurement of Correlation Between Unresolved IRAS Sources and Optically-Selected Galaxy Clusters}

\author[a]{Adam D. Hincks,}
\emailAdd{hincks@cita.utoronto.ca}
\author[a]{Amir Hajian}
\emailAdd{ahajian@cita.utoronto.ca}
\author[b]{and Graeme E. Addison}
\emailAdd{gaddison@phas.ubc.ca}

\affiliation[a]{Canadian Institute for Theoretical Astrophysics, University of Toronto,\hspace{1em}Toronto, ON M5S~3H8, Canada}
\affiliation[b]{Department of Physics and Astronomy, University of British Columbia, Vancouver, BC V6T~1Z4, Canada}

\abstract{We cross-correlate the $100\,\micro\metre$ Improved Reprocessing of the IRAS Survey (IRIS) map and galaxy clusters at $0.1<z<0.3$ in the maxBCG catalogue taken from the Sloan Digital Sky Survey, measuring an angular cross-power spectrum over multipole moments $150<\ell<3000$ at a total significance of over $40\sigma$. The cross-spectrum, which arises from the spatial correlation between unresolved dusty galaxies that make up the cosmic infrared background (CIB) in the IRIS map and the galaxy clusters, is well-fit by a single power law with an index of $-1.28\pm 0.12$, similar to the clustering of unresolved galaxies from cross-correlating far-infrared and submillimetre maps at longer wavelengths. Using a recent, phenomenological model for the spectral and clustering properties of the IRIS galaxies, we constrain the large-scale bias of the maxBCG clusters to be $2.6\pm1.4$, consistent with existing analyses of the real-space cluster correlation function. The success of our method suggests that future CIB--optical cross-correlations using \emph{Planck} and \emph{Herschel} data will significantly improve our understanding of the clustering and redshift
distribution of the faint CIB sources.}

\maketitle
\flushbottom

\section{Introduction}
\label{sec:intro}

The cosmic infrared background (CIB), which peaks at $\sim200\,\micro\metre$, largely arises from thermal re-emission of UV and optical starlight by dust grains, and comprises nearly half of all extragalactic radiation \cite{puget/etal:1996, fixsen/etal:1998, lagache/puget:2000}. Because ground-based observations suffer from atmospheric absorption and balloon/space-based telescopes have had limited resolution, characterising the far-infrared and submillimetre (submm) CIB using individual sources---predominantly dusty, star-forming galaxies (DSFGs) at $0<z<4$---has only been possible fairly recently by using statistical techniques, such as stacking (e.g., \cite{dole/etal:2006, devlin/etal:2009, marsden/etal:2009, pascale/etal:2009,bethermin/etal:2012}).

Given the difficulty in resolving CIB sources at long wavelengths, studying fluctuations in the CIB (e.g., \cite{bond/carr/hogan:1986,bond/carr/hogan:1991,haiman/knox:2000}) has emerged as a complementary method to constrain its properties. Clustering of unresolved CIB sources was first detected at $160\,\micro\metre$ in Spitzer data \cite{lagache/etal:2007, grossan/smoot:2007}, and has since been measured by the Balloon-borne Large Aperture Submillimeter Telescope (BLAST; \cite{viero/etal:2009,hajian/etal:2012}), \emph{Herschel's} Spectral and Photometric Imaging Receiver (SPIRE; e.g., \cite{amblard/etal:2011,viero/etal:prep}) and \emph{Planck}'s High Frequency Instrument \cite{planck:2011}. The South Pole Telescope \cite{hall/etal:2010} detected clustering of dusty galaxies at millimetre wavelengths; shortly thereafter, Ref.~\cite{hajian/etal:2012} measured the clustering of the CIB by cross-correlating submm BLAST maps with millimetre Atacama Cosmology Telescope maps.

The clustering of unresolved CIB sources has been used to infer their bias relative to the underlying matter fluctuations and to characterise the dark matter halos with which they are associated (e.g., \cite{viero/etal:2009,amblard/etal:2011,planck:2011,shang/etal:2011,xia/etal:2012}). However, extracting clustering properties from angular power spectra of CIB anisotropies is challenging because of Galactic dust emission on large angular scales and shot-noise from auto-correlation of unresolved sources on small angular scales. The redshift-distribution of the unresolved CIB sources also remains uncertain, particularly at wavelengths longward of $500\,\micro\metre$.

One way to reduce or bypass these issues is to cross-correlate maps containing unresolved CIB sources with a catalogue of extragalactic sources with known redshifts. We would not expect Galactic emission to be correlated with such a catalogue, so while Galactic dust exists as effective noise in the CIB map, it would not add to the cross-correlation signal. A shot-noise term on small scales would indicate the extent to which CIB emission is associated with the catalogue members, which could be of interest in its own right.

In this paper, we present such an analysis by cross-correlating the 100$\micro\metre$ Improved Reprocessing of the IRAS Survey (IRIS) map \cite{mivilledeschenes/lagache:2005} with the positions of optically-selected galaxy clusters in the maxBCG catalogue \cite{koester/etal:2007a}. Clustering of the DSFGs in IRIS map has been studied before. Ref.~\cite{hajian/etal:2012} marginally detected a weak clustering term in the IRIS power spectrum of a $\sim\!9\,\degword^2$ field with minimum emission from the Galactic cirrus near the South Ecliptic Pole. Later, Ref.~\cite{penin/etal:2012} measured the CIB anisotropies in IRIS maps that were cleaned of Galactic cirrus using cross-correlation with HI maps. Our measurement is not sensitive to assumptions about Galactic contamination, which acts only as additional noise, as mentioned above, and is not dependent on any external tracer of the Galactic emission; the only significant contribution to the cross-spectrum signal comes from the spatial correlation of the IRIS DSFGs and the maxBCG clusters.

The spatial clustering of DSFGs and galaxy clusters is also relevant for ongoing measurements of the kinematic Sunyaev-Zel'dovich (kSZ; \cite{sunyaev/zeldovich:1980}) effect, i.e., the Doppler shift of cosmic microwave background (CMB) photons by electrons possessing some bulk motion in the presence of a density or ionisation inhomogeneity. Measuring the kSZ angular power spectrum is of great interest for constraining an extended, `patchy' reionisation scenario (e.g., \cite{zahn/etal:2012,battaglia/etal:prep,park/etal:prep}), but current constraints are degraded because of uncertainty in the degree of correlation between the CIB sources and the thermal Sunyaev Zel'dovich effect (tSZ; \cite{sunyaev/zeldovich:1972}) from the scattering of CMB photons off energetic electrons in the potential wells of galaxy groups and clusters (e.g., \cite{shirokoff/etal:2011,reichardt/etal:2012,mesinger/mcquinn/spergel:2012,sievers/etal:prep}).

This paper is organised as follows. In Section~\ref{sec:reduction}, we describe our measurement of the IRIS$\times$maxBCG cross-spectrum and our uncertainty estimates; then, in Section~\ref{sec:model}, we describe and discuss a simple model that we fit to the measured spectrum. A discussion and conclusion follow in Sections~\ref{sec:discussion} and \ref{sec:conclusions}.

\section{Data Reduction}
\label{sec:reduction}

\subsection{Data Sets, Maps and Masks}
\label{ssec:data_sets}

Here, we describe the two data sets used in this paper: the IRIS $100\,\micro\meter$ map and the maxBCG catalogue.

The IRIS maps are a reprocessed version of the Infrared Astronomical Satellite (IRAS) survey \cite{beichman/etal:1988} which mapped 98\% of the sky at $12$, $25$, $60$ and $100\,\micro\metre$. Though IRAS was originally designed for discovering and characterising compact sources, it was increasingly found useful for studying diffuse, infrared emission. To aid in the latter, particularly with the goal of having suitable maps for foreground cleaning of the upcoming \emph{Planck} satellite, IRIS reprocessed the IRAS plates, significantly improving large-scale calibration, destriping and zodiacal light removal. In this paper, we use their publicly-available full-sky, $100\,\micro\metre$, `no hole' map with HEALPix \cite{gorski/etal:2005} resolution of {\tt NSIDE} = 1024.\footnote{\url{http://www.cita.utoronto.ca/~mamd/IRIS/IrisDownload.html}} It has a noise level of $(0.06 \pm 0.02)\,\mega\jansky\,\steradian^{-1}$ and an effective resolution of $(4.3 \pm 0.2)\,\mathrm{arcmin}$, with the latter obtained by assuming that the effective point spread function was a Gaussian. The overall calibration uncertainty is estimated at 13.5\% and is limited by the uncertainty of the Diffuse Infrared Background Experiment (DIRBE) maps which were used for calibration. More details on the IRIS maps are in Ref.~\cite{mivilledeschenes/lagache:2005}.

\begin{figure}[t]
  \center\input{dist}
  \caption{Comparison of the redshift distributions of the maxBCG cluster catalogue and the unresolved IRIS galaxies. The solid red histogram shows the distribution of maxBCG clusters, $dN / dz$ (with $dz = 0.01$), where redshifts are photometric and have errors $\leq 3\%$ \cite{koester/etal:2007a}. The IRIS curve (blue line) shows the flux-weighted redshift distribution, $dI/dz$, of IRIS sources from the CIB model of Ref.~\cite{addison/dunkley/bond:prep}, used for the fitting described in Section~\ref{ssec:cluster_bias_model}. Each distribution has been normalised so that the area under it is unity.}
  \label{fig:z_dist}
\end{figure}
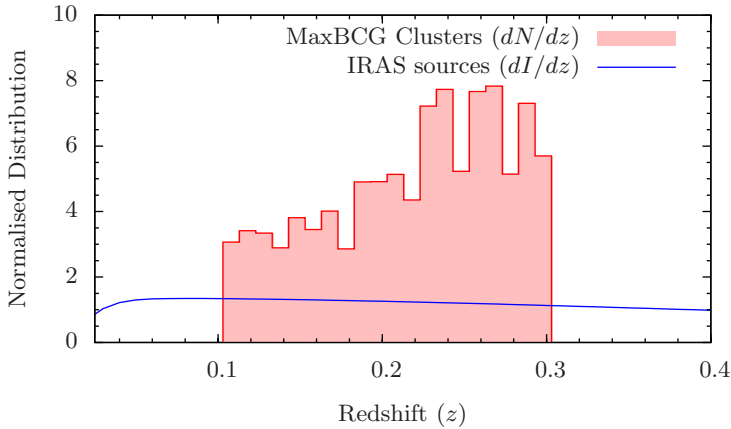

The maxBCG catalogue of galaxy clusters was compiled by running the maxBCG red-sequence selection algorithm \cite{koester/etal:2007a} on the Sloan Digital Sky Survey (SDSS) photometric data \cite{york/etal:2000}. The catalogue consists of 13,823 clusters of galaxies of richness $10 \leq N_{\mathrm{gal}}^{R_{200}} \leq 190$ over redshifts $0.1 < z < 0.3$ in an area about $7500\,\degword^2$, and is estimated to be more than 90\% pure and more than 90\% complete for clusters more massive than $2\times10^{14} h^{-1} M_{\odot}$.\footnote{$N_{\mathrm{gal}}^{R_{200}}$ is the number of galaxies within a radius containing a galaxy density which is 200 times larger than the average density.} Fig.~\ref{fig:z_dist} shows the redshift distribution of the clusters, compared to the IRIS source distribution used in the model described in Section~\ref{ssec:cluster_bias_model}. See Ref.~\cite{koester/etal:2007a} for further details on the maxBCG catalogue.

In our work, we construct a maxBCG overdensity map by creating a blank map of zeros at the same resolution as the IRIS map ({\tt NSIDE} = 1024), and then setting to unity any pixel containing a cluster coordinate from the maxBCG catalogue. The mean is then removed and the map scaled by the inverse number of illuminated pixels, i.e.:

\begin{equation}
  \bm{m}' = \frac{\bm{m} - \langle m \rangle}{\langle m \rangle},
\end{equation}

\noindent where $\bm{m}$ is the map of ones and zeros and $\langle m \rangle$ is the mean value of $\bm{m}$ after applying the maxBCG mask (see below).

\begin{figure}[t]
  \center\input{masked}
  \caption{The masked, full sky IRIS map, centred on the Galactic north pole with Mollweide's projection. The figure shows both the masks---the IRIS mask, to remove high Galactic emission and point sources, and the maxBCG mask, selecting the survey area (which is mainly in northern Galactic latitudes). The finger-like portion of the mask near the north pole is to remove a flaw in the IRIS map. Total coverage is 6.41\%, or $2644\,\degword^2$.}
  \label{fig:maps}
\end{figure}
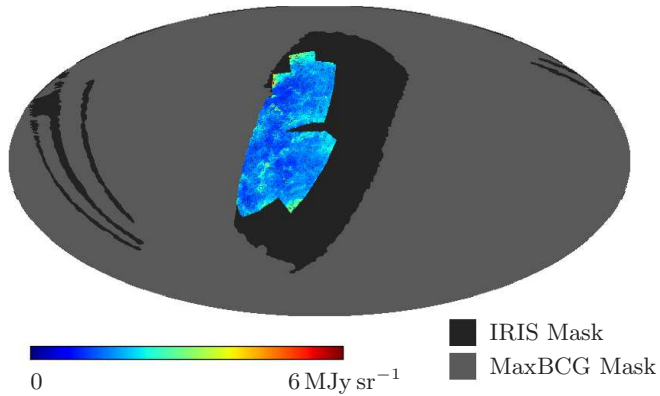

Each of the IRIS and maxBCG maps are masked when computing the cross-spectrum (Sec.~\ref{ssec:xspec_anal}). For maxBCG, we simply mask all regions falling outside the maxBCG survey. The IRIS mask excludes (a) regions with high Galactic infrared emission and (b) resolved point sources.  For (a), we do a conservative cut and mask all IRIS tiles which have a mean map value more than $2.0\,\mega\jansky\,\steradian^{-1}$. This is to minimise potential noise due to Galactic cirrus emission. Additionally, we mask a finger-shaped region near the north pole where the IRIS map is of particularly low quality. For component (b), we mask all entries in the IRAS Point Source Catalog \cite{beichman/etal:1988} and the IRAS Small-Scale Structure Catalog \cite{helou/walker:1988} with discs of radius $5'$. The objects in these IRAS catalogues are typically bright ($\gtrsim 1\,\jansky$), very nearby sources that we do not expect to add to our cross-correlation signal. The sky coverage after both IRIS and maxBCG masks are applied is 6.41\%, or $2644\,\degword^2$, and contains 4784 clusters. Figure~\ref{fig:maps} shows the IRIS $100\,\micro\metre$ map with each of the masks.

\subsection{Computation of the Cross-Spectrum}
\label{ssec:xspec_anal}

\begin{figure}[t]
  \center\input{cross_spectrum}
  \caption{IRIS$\times$maxBCG cross-spectrum. The cross-spectrum bandpowers are shown in the upper panel along with a power-law fit with $C_{\ell}\propto\ell^{-1.28}$, as well as the best fit that uses the phenomenological IRIS DSFG model from Ref.~\cite{addison/dunkley/bond:prep}, with a parametrisation of the cluster bias (see Section~\ref{ssec:cluster_bias_model}). The lower panel shows a null test taken by cross-correlating random-realisations of maps having the IRIS autospectrum with the maxBCG map; note that this plot has a linear scale on the $y$-axis. The result is consistent with zero power.}
  \label{fig:cross_spectra}
\end{figure}
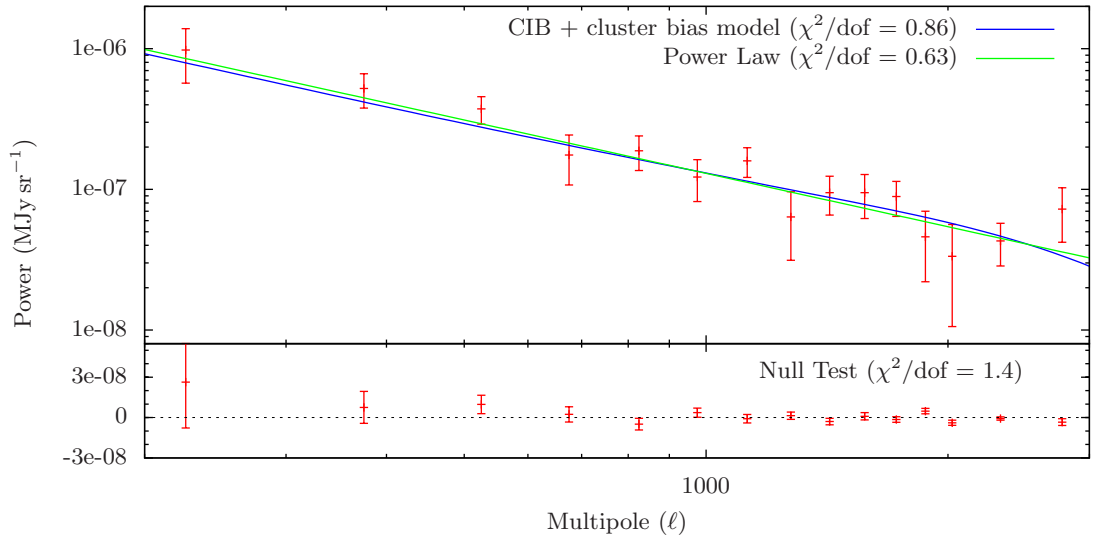

The cross-spectrum and autospectra are calculated using {\tt PolSpice} \cite{chon/etal:2004}. We correct for a Gaussian beam with FWHM of $4.3\arcminute$ \cite{mivilledeschenes/lagache:2005} as well as for a residual transfer function that is left over by {\tt PolSpice} due to imperfect correction for the $5'$ point source mask. This transfer function, which is small compared to the uncertainty in our power spectrum measurement, was determined by measuring the power spectrum of a synthetic map of white noise after applying the same mask. Finally, we bin the cross-spectrum into bands of width $\Delta\ell = 150$ below $\ell = 2000$; at higher multipoles, where the data are noisier, we choose a bin width of $\Delta\ell = 450$. The cross-spectrum is shown in Fig.~\ref{fig:cross_spectra}.

To estimate our uncertainties, we generate 144 random realisations of maps with the same power spectrum as the IRIS map, and then cross-correlate this map with the cluster map in the same way as described above. Their average spectrum, consisting of the mean of these 144 synthetic spectra in each $\Delta\ell$-band, is consistent with a null-measurement, having $\chi^2$ = 18.4 for 13 degrees of freedom (dof), with a probability to exceed (PTE) of 0.14. This null-test is shown in Fig.~\ref{fig:cross_spectra}. We take the covariance of these mean band powers as our estimate of the measured spectrum uncertainties.

\begin{figure}[t]
  \center\input{vary_masks}
  \caption{Tests for systematic errors due to the choice of masks (c.f., Table~\ref{tab:mask_cuts}), which show that our mask choice is not biasing the cross-spectrum. {\em Left:} The mean intensity cut for allowing IRIS tiles into the unmasked region is varied from 1.65 to $3.0\,\mega\jansky\,\steradian^{-1}$. {\em Right:} The radius of point-source masks is varied from 2 to $10\arcminute$.}
  \label{fig:compare_mask}
\end{figure}
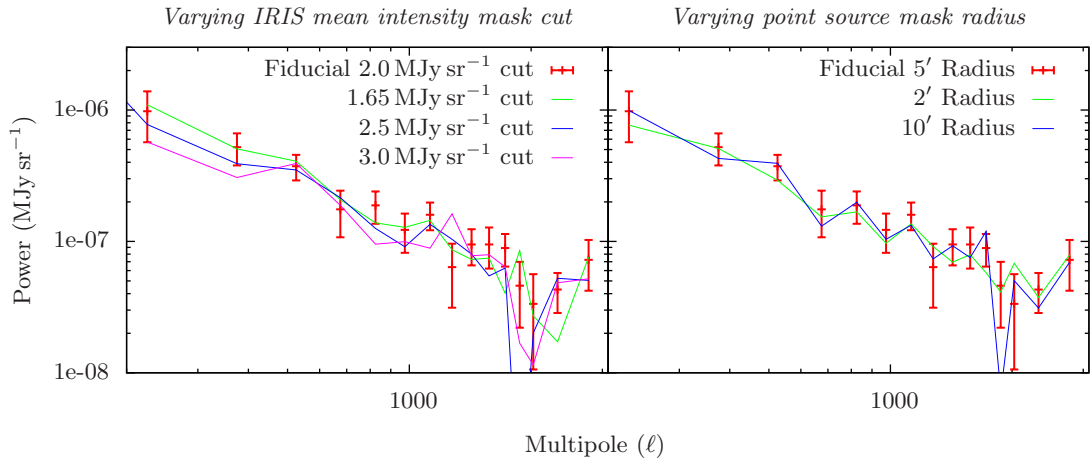

The choice to mask IRIS panels with mean intensity greater than $2.0\,\mega\jansky\,\steradian^{-1}$ is conservative, as is the $5\arcminute$-radius masking of IRAS point and compact sources. Nevertheless, in order to verify that the masks do not bias our result (which could potentially occur if, for example, fewer maxBCG clusters are detected where the Galactic emission is brightest), we computed power spectra with different mask sizes. We varied both the mask size based on the IRIS mean intensity, as well as the radius of point-source masks. Table~\ref{tab:mask_cuts} lists all the mask combinations that we performed along with their respective sky-coverage. Figure~\ref{fig:compare_mask} shows that the different mask sizes do not significantly alter the cross-spectrum. In the end, the $2.0\,\mega\jansky\,\steradian^{-1}$ IRIS threshold plus $5\arcminute$ point source mask combination was chosen as it balanced sky area against low cross-spectrum uncertainties.

\section{Model fitting}
\label{sec:model}

\subsection{Power law fit}

We first model the IRIS$\times$maxBCG cross-spectrum as a power law of the form $\mathcal{C}_{\ell}=A\,(\ell/1000)^{\gamma}$. This simple model fits the data well, with $\chi^2$ per dof = $8.2/13$ (PTE = 0.83). We measure $A=(1.3\pm0.2)\times10^{-7}\,\mega\jansky\,\steradian^{-1}$ and $\gamma=-1.28\pm0.12$, where the uncertainty in $A$ includes, and is dominated by, the 13.5\% uncertainty in the absolute flux calibration of the DIRBE maps used to re-calibrate the original IRAS maps by Ref.~\cite{mivilledeschenes/lagache:2005}.

A model assuming no correlation between the IRIS and maxBCG overdensity maps, where the cross-spectrum is zero at all scales, yields a $\chi^2$ of 221. Comparing to the power-law fit, we infer that IRIS$\times$maxBCG cross-correlation is detected at a significance level of over 40$\sigma$.

\subsection{Large-scale cluster bias}
\label{ssec:cluster_bias_model}

We also use our cross-spectrum measurement to constrain the large-scale bias of the maxBCG clusters, in order to compare with previous clustering analyses using the maxBCG catalogue. To do this, we use an existing model for the properties of the unresolved dusty galaxies in the IRIS map. Ref.~\cite{addison/dunkley/bond:prep} (hereafter A12) present a phenomenological model for the luminosity function, spectral energy distribution and clustering properties of CIB sources, constrained by differential number counts and angular power spectra from $70\,\micro\metre$ to $1.4\,$mm. We write the IRIS$\times$maxBCG cross-spectrum at multipole $\ell$ as (following, e.g., Refs.~\cite{haiman/knox:2000,ho/etal:2008}):
\begin{equation}
\mathcal{C}_{\ell}=\int \frac{dz}{\chi^2}\frac{1}{(1+z)}\langle b_{\rm gal}(k,z)\rangle\langle b_{\rm clust}(k,z)\rangle\,\langle j_{\nu}(z)\rangle_{\rm cut}\frac{dN_{\rm clust}/dz}{\int dz'\,dN_{\rm clust}/dz'} P_{\rm DM}(k,z)|_{k=\frac{\ell+1/2}{\chi}},
\end{equation}
where $\chi$ is the comoving distance to redshift $z$, $b_{\rm gal}$ is the mean IRIS source bias, $b_{\rm clust}$ is the mean cluster bias, $j_{\nu}$ is the IRIS source emissivity density, $dN_{\rm clust}/dz$ is the redshift-distribution of the maxBCG clusters and $P_{\rm DM}$ is the linear matter power spectrum. Since our spectra are from scales $\ell>100$, we apply the small-scale Limber approximation \citep{limber:1953,kaiser:1992}, setting $k=(\ell+1/2)/\chi$. The subscript `cut' denotes the fact that bright sources ($S_{100\,\micro\metre}\gtrsim1$~Jy) have been masked in the IRIS map; expressions for $\langle b_{\rm gal}\rangle$ and $\langle j_{\nu}\rangle_{\rm cut}$ are given in Section~2 of A12. Fig.~\ref{fig:z_dist} shows a comparison of the redshift distributions of the model's IRIS sources and the maxBCG galaxy clusters.
\begin{table}[t]
  \center
  \begin{tabular}{cc|cccc}
      & & \multicolumn{4}{c}{{\em IRIS Intensity Threshold (MJy\,sr$\mathit{^{-\!1}}\!$)}} \\
      & & 1.65 & {\bf 2.0} & 2.5 & 3.0 \\
    \hline
    \multirow{3}{*}{\begin{tabular}{c}{\em Point Source}\\
                                      {\em Mask (arcmin)}\end{tabular}} &
      2 & -- & 6.57\% & -- & -- \\
    & {\bf 5} & 4.59\% & {\bf 6.41}\% & 10.34\% & 14.46\% \\
    & 10 & -- & 6.00\% & -- & -- \\
  \end{tabular}
  \caption{Sky coverage of masks used to test for systematics. The overall mask was varied by admitting different thresholds of IRIS average intensity (columns), while the point-source mask was varied in radius (rows). The fiducial mask, used for all our analysis, is the central entry in bold type.}
  \label{tab:mask_cuts}
\end{table}

We repeated the fit described in A12 with the addition of the IRIS$\times$maxBCG cross-spectrum. We found that the cross-spectrum cannot be described using a scale-independent, linear, cluster bias; this is not surprising since over the range of physical scales probed by our measurement ($0.15\lesssim k/\textrm{Mpc}^{-1}\lesssim7.5$), the matter power spectrum becomes strongly non-linear. Part, or all, of the non-linear clustering may arise from IRIS sources that are actually members of the maxBCG clusters (the `one-halo' term of the halo model clustering formalism; for a review see Ref.~\cite{cooray/sheth:2002}). In the present analysis, we do not attempt to separately model this term, and instead simply parametrize the cluster bias relative to the linear matter power spectrum using a simple phenomenological expansion in powers of $k$:
\begin{equation}
\langle b_{\rm clust}(k,z)\rangle=b_0\left[1+A_{\rm bias}\left(\frac{k}{k_0}\right)+\mathcal{O}(k^2)\right],
\end{equation}
with two free parameters, the large-scale bias, $b_0$, and a parameter describing the scale dependence, $A_{\rm bias}$. The pivot wavenumber $k_0$ is chosen to be 1 Mpc$^{-1}$. We find that the data are unable to constrain higher-order terms in this expansion.

Compared to the original A12 fit, the $\chi^2$ increases by $\Delta\chi^2=11.2$ for 13 extra degrees of freedom (15 IRIS$\times$maxBCG bandpowers minus two new parameters, $b_0$ and $A_{\rm bias}$). This corresponds to a PTE of 0.59, indicating that the IRIS$\times$maxBCG cross-spectrum is well-described by the A12 CIB source model with the addition of the cluster bias parameters. None of the A12 model parameters are significantly shifted compared to the fit without the IRIS$\times$maxBCG cross-spectrum. The large-scale maxBCG cluster bias is constrained to be $b_0=2.6\pm1.4$, which agrees with the measurement of $2.80\pm0.13$ from Ref.~\cite{estrada/sefusatti/frieman:2009}, obtained using the real-space maxBCG cluster correlation function. The nuisance parameter $A_{\rm bias}$ is measured to be $3.9\pm2.5$, and is highly anti-correlated with $b_0$ (see below). The best-fit IRIS$\times$maxBCG model is plotted in the top panel of Figure~\ref{fig:cross_spectra}.

\section{Discussion}
\label{sec:discussion}

While we have measured a strong cross-correlation signal, our constraint on the cluster bias is weak compared to the real-space maxBCG correlation function measurement of Ref.~\cite{estrada/sefusatti/frieman:2009}. This is primarily because, as stated, our measurement probes non-linear scales, and so is unable to distinguish the large-scale cluster bias from non-linear effects, leading to a strong degeneracy between $b_0$ and $A_{\rm bias}$. Extracting the quasi-linear and linear-scale correlation ($k\lesssim0.1~$Mpc$^{-1}$, corresponding to $\ell\lesssim100$ for this redshift range) using IRIS data is difficult because of effective noise from Galactic dust emission.  We expect the imminent release of submm and microwave \emph{Planck} maps to lead to significant improvements in this measurement. While the \emph{Planck} resolution is comparable to IRIS, having multiple frequency channels will both facilitate reducing noise from Galactic dust and provide a stronger test of CIB source models. A joint analysis of the large-scale cluster correlation function and our cross-spectrum measurement, with an appropriate treatment of the covariance, would also help to disambiguate the linear and non-linear clustering.

It is intriguing that the shape of the IRIS$\times$maxBCG cross-spectrum is similar to that of the clustered dusty galaxy component measured from existing FIR, submm and microwave measurements ($C_{\ell}\propto\ell^{\sim-1.2}$, e.g., Ref.~\cite{amblard/etal:2011,addison/etal:2012,reichardt/etal:2012}). Future work is clearly required to understand the implications of the apparent similarity, and the extent to which the small-scale CIB--cluster cross-correlations can be used to tell us about the occupation of clusters with dusty galaxies or how star-forming regions are distributed within clusters. The \emph{Herschel} SPIRE instrument has channels at 250, 350 and $500\,\micro\metre$ and roughly ten times better resolution than IRIS or \emph{Planck}, and should provide improved constraints on CIB--cluster cross-correlations on smaller angular scales. SPIRE sky coverage includes over a hundred square degrees overlap with SDSS and upcoming optical surveys.\footnote{\url{http://www.astro.caltech.edu/~viero/viero_homepage/helms_hers.html}}

The significant correlation between the IRIS map and the maxBCG clusters qualitatively supports the prediction of a moderate degree of spatial correlation between the thermal Sunyaev Zel'dovich effect associated with groups and clusters and faint dusty galaxies that make up the CIB \cite{addison/dunkley/spergel:2012}, relevant for current efforts to constrain the kinetic Sunyaev Zel'dovich effect \citep{reichardt/etal:2012}. We note, however, that the maxBCG clusters are all at $z<0.3$, whereas most of the tSZ power comes from groups and clusters at higher redshift (see, e.g., Figure 1 of Ref.~\cite{shaw/etal:2009}). Nevertheless, our result suggests that useful quantitative constraints on the tSZ--CIB correlation may be obtained by cross-correlation of catalogues of galaxy clusters at higher redshifts with FIR, submm and microwave maps.

We have not accounted for any spatial offset between BCG position and cluster centre in our analysis. Ref.~\cite{zitrin/etal:2012} find that such offsets mostly lie in the range $-2.5<\textrm{log}(\Delta r/h^{-1}\textrm{Mpc})<-1.0$, where $\Delta r$ is the physical offset of the BCG from the cluster centre of mass, using strong gravitational lensing. While such offsets would lead to a suppression of small-scale power in the IRIS$\times$maxBCG cross-spectrum, their effect is small for $\ell<3000$. We similarly have not accounted for impurity in the maxBCG catalogue. Clearly, these uncertainties may be expected to become important for analyses attempting a more physical interpretation of CIB--cluster cross-correlations.

\section{Conclusions}
\label{sec:conclusions}

We have measured the cross-power spectrum of the $100\,\micro\metre$ IRIS map and an overdensity map constructed from the maxBCG catalogue of optically-selected clusters at $0.1<z<0.3$. The cross-spectrum is detected at a statistical significance of over $40\sigma$. This result is possible because the dominant components of the IRIS auto-correlation power spectrum---Galactic dust on large scales and Poisson-like power from self-correlation of DSFGs on small scales---are not present in the cross-spectrum signal, allowing information about the spatial correlation of the faint IRIS sources and maxBCG clusters to be cleanly extracted.

While our result provides only a fairly weak test of the A12 model of CIB sources, it has demonstrated the potential of CIB--optical cross-correlations. We anticipate rapid improvements in this field with upcoming analyses using \emph{Planck} and \emph{Herschel} data. Looking beyond our analysis, which has focussed on galaxy clusters, a wealth of information may be gleaned using other optically-selected populations (e.g., quasars and luminous red galaxies). CIB--Quasar cross-correlations, for example, may help elucidate the connection between galactic nuclei activity and star formation, and how this relationship evolves with redshift. The use of large catalogues of objects with well-measured redshifts in cross-correlations will also provide important constraints on the redshift-distribution of unresolved CIB sources.

\acknowledgments
GA acknowledges support from a CITA National Fellowship. Some of the results in this paper have been derived using the HEALPix package \cite{gorski/etal:2005}. Cross-spectra were calculated using {\tt PolSpice} \cite{chon/etal:2004}.

\bibliography{ixc}
\bibliographystyle{JHEP}

\end{document}

%% file: dist.tex
\begingroup
  \footnotesize
  \selectfont
\makeatletter
\providecommand\color[2][]{%
\GenericError{(gnuplot) \space\space\space\@spaces}{%
Package color not loaded in conjunction with
terminal option `colourtext'%
}{See the gnuplot documentation for explanation.%
}{Either use 'blacktext' in gnuplot or load the package
color.sty in LaTeX.}%
\renewcommand\color[2][]{}%
}%
\providecommand\includegraphics[2][]{%
\GenericError{(gnuplot) \space\space\space\@spaces}{%
Package graphicx or graphics not loaded%
}{See the gnuplot documentation for explanation.%
}{The gnuplot epslatex terminal needs graphicx.sty or graphics.sty.}%
\renewcommand\includegraphics[2][]{}%
}%
\providecommand\rotatebox[2]{#2}%
\@ifundefined{ifGPcolor}{%
\newif\ifGPcolor
\GPcolortrue
}{}%
\@ifundefined{ifGPblacktext}{%
\newif\ifGPblacktext
\GPblacktexttrue
}{}%
\let\gplgaddtomacro\g@addto@macro
\gdef\gplbacktext{}%
\gdef\gplfronttext{}%
\makeatother
\ifGPblacktext
\def\colorrgb#1{}%
\def\colorgray#1{}%
\else
\ifGPcolor
\def\colorrgb#1{\color[rgb]{#1}}%
\def\colorgray#1{\color[gray]{#1}}%
\expandafter\def\csname LTw\endcsname{\color{white}}%
\expandafter\def\csname LTb\endcsname{\color{black}}%
\expandafter\def\csname LTa\endcsname{\color{black}}%
\expandafter\def\csname LT0\endcsname{\color[rgb]{1,0,0}}%
\expandafter\def\csname LT1\endcsname{\color[rgb]{0,1,0}}%
\expandafter\def\csname LT2\endcsname{\color[rgb]{0,0,1}}%
\expandafter\def\csname LT3\endcsname{\color[rgb]{1,0,1}}%
\expandafter\def\csname LT4\endcsname{\color[rgb]{0,1,1}}%
\expandafter\def\csname LT5\endcsname{\color[rgb]{1,1,0}}%
\expandafter\def\csname LT6\endcsname{\color[rgb]{0,0,0}}%
\expandafter\def\csname LT7\endcsname{\color[rgb]{1,0.3,0}}%
\expandafter\def\csname LT8\endcsname{\color[rgb]{0.5,0.5,0.5}}%
\else
\def\colorrgb#1{\color{black}}%
\def\colorgray#1{\color[gray]{#1}}%
\expandafter\def\csname LTw\endcsname{\color{white}}%
\expandafter\def\csname LTb\endcsname{\color{black}}%
\expandafter\def\csname LTa\endcsname{\color{black}}%
\expandafter\def\csname LT0\endcsname{\color{black}}%
\expandafter\def\csname LT1\endcsname{\color{black}}%
\expandafter\def\csname LT2\endcsname{\color{black}}%
\expandafter\def\csname LT3\endcsname{\color{black}}%
\expandafter\def\csname LT4\endcsname{\color{black}}%
\expandafter\def\csname LT5\endcsname{\color{black}}%
\expandafter\def\csname LT6\endcsname{\color{black}}%
\expandafter\def\csname LT7\endcsname{\color{black}}%
\expandafter\def\csname LT8\endcsname{\color{black}}%
\fi
\fi
\setlength{\unitlength}{0.0500bp}%
\begin{picture}(5040.00,3240.00)%
\gplgaddtomacro\gplbacktext{%
\csname LTb\endcsname%
\put(-374,1674){\rotatebox{-270}{\makebox(0,0){\strut{}Normalised Distribution}}}%
\put(2519,-110){\makebox(0,0){\strut{}Redshift ($z$)}}%
}%
\gplgaddtomacro\gplfronttext{%
\csname LTb\endcsname%
\put(3854,2736){\makebox(0,0)[r]{\strut{}MaxBCG Clusters ($dN/dz$)}}%
\csname LTb\endcsname%
\put(3854,2516){\makebox(0,0)[r]{\strut{}IRAS sources ($dI/dz$)}}%
\csname LTb\endcsname%
\put(66,440){\makebox(0,0)[r]{\strut{} 0}}%
\put(66,934){\makebox(0,0)[r]{\strut{} 2}}%
\put(66,1428){\makebox(0,0)[r]{\strut{} 4}}%
\put(66,1921){\makebox(0,0)[r]{\strut{} 6}}%
\put(66,2415){\makebox(0,0)[r]{\strut{} 8}}%
\put(66,2909){\makebox(0,0)[r]{\strut{} 10}}%
\put(1127,220){\makebox(0,0){\strut{} 0.1}}%
\put(2365,220){\makebox(0,0){\strut{} 0.2}}%
\put(3603,220){\makebox(0,0){\strut{} 0.3}}%
\put(4841,220){\makebox(0,0){\strut{} 0.4}}%
}%
\gplbacktext
\put(0,0){\includegraphics{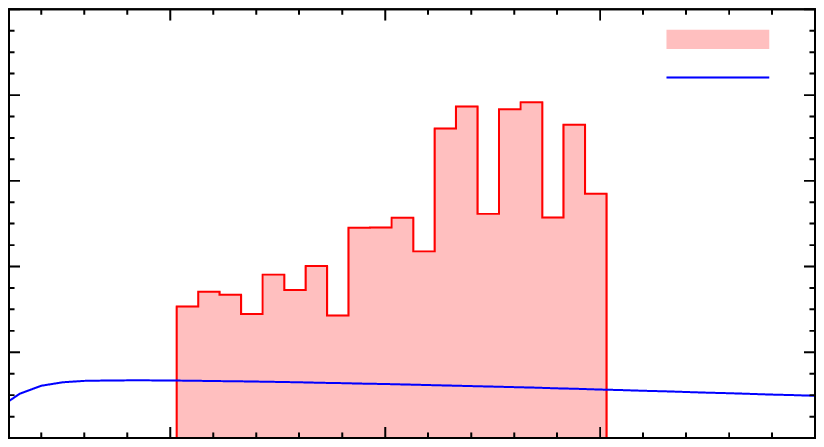}}%
\gplfronttext
\end{picture}%
\endgroup

%% file: masked.tex
\begingroup
  \footnotesize
  \selectfont
\makeatletter
\providecommand\color[2][]{%
\GenericError{(gnuplot) \space\space\space\@spaces}{%
Package color not loaded in conjunction with
terminal option `colourtext'%
}{See the gnuplot documentation for explanation.%
}{Either use 'blacktext' in gnuplot or load the package
color.sty in LaTeX.}%
\renewcommand\color[2][]{}%
}%
\providecommand\includegraphics[2][]{%
\GenericError{(gnuplot) \space\space\space\@spaces}{%
Package graphicx or graphics not loaded%
}{See the gnuplot documentation for explanation.%
}{The gnuplot epslatex terminal needs graphicx.sty or graphics.sty.}%
\renewcommand\includegraphics[2][]{}%
}%
\providecommand\rotatebox[2]{#2}%
\@ifundefined{ifGPcolor}{%
\newif\ifGPcolor
\GPcolortrue
}{}%
\@ifundefined{ifGPblacktext}{%
\newif\ifGPblacktext
\GPblacktexttrue
}{}%
\let\gplgaddtomacro\g@addto@macro
\gdef\gplbacktext{}%
\gdef\gplfronttext{}%
\makeatother
\ifGPblacktext
\def\colorrgb#1{}%
\def\colorgray#1{}%
\else
\ifGPcolor
\def\colorrgb#1{\color[rgb]{#1}}%
\def\colorgray#1{\color[gray]{#1}}%
\expandafter\def\csname LTw\endcsname{\color{white}}%
\expandafter\def\csname LTb\endcsname{\color{black}}%
\expandafter\def\csname LTa\endcsname{\color{black}}%
\expandafter\def\csname LT0\endcsname{\color[rgb]{1,0,0}}%
\expandafter\def\csname LT1\endcsname{\color[rgb]{0,1,0}}%
\expandafter\def\csname LT2\endcsname{\color[rgb]{0,0,1}}%
\expandafter\def\csname LT3\endcsname{\color[rgb]{1,0,1}}%
\expandafter\def\csname LT4\endcsname{\color[rgb]{0,1,1}}%
\expandafter\def\csname LT5\endcsname{\color[rgb]{1,1,0}}%
\expandafter\def\csname LT6\endcsname{\color[rgb]{0,0,0}}%
\expandafter\def\csname LT7\endcsname{\color[rgb]{1,0.3,0}}%
\expandafter\def\csname LT8\endcsname{\color[rgb]{0.5,0.5,0.5}}%
\else
\def\colorrgb#1{\color{black}}%
\def\colorgray#1{\color[gray]{#1}}%
\expandafter\def\csname LTw\endcsname{\color{white}}%
\expandafter\def\csname LTb\endcsname{\color{black}}%
\expandafter\def\csname LTa\endcsname{\color{black}}%
\expandafter\def\csname LT0\endcsname{\color{black}}%
\expandafter\def\csname LT1\endcsname{\color{black}}%
\expandafter\def\csname LT2\endcsname{\color{black}}%
\expandafter\def\csname LT3\endcsname{\color{black}}%
\expandafter\def\csname LT4\endcsname{\color{black}}%
\expandafter\def\csname LT5\endcsname{\color{black}}%
\expandafter\def\csname LT6\endcsname{\color{black}}%
\expandafter\def\csname LT7\endcsname{\color{black}}%
\expandafter\def\csname LT8\endcsname{\color{black}}%
\fi
\fi
\setlength{\unitlength}{0.0500bp}%
\begin{picture}(4680.00,3168.00)%
\gplgaddtomacro\gplbacktext{%
}%
\gplgaddtomacro\gplfronttext{%
\csname LTb\endcsname%
\put(3626,554){\makebox(0,0)[l]{\strut{}IRIS Mask}}%
\put(3626,301){\makebox(0,0)[l]{\strut{}MaxBCG Mask}}%
\put(164,190){\makebox(0,0)[l]{\strut{}0}}%
\put(2106,190){\makebox(0,0)[l]{\strut{}$6\,\mega\jansky\,\steradian^{-1}$}}%
}%
\gplbacktext
\put(0,0){\includegraphics{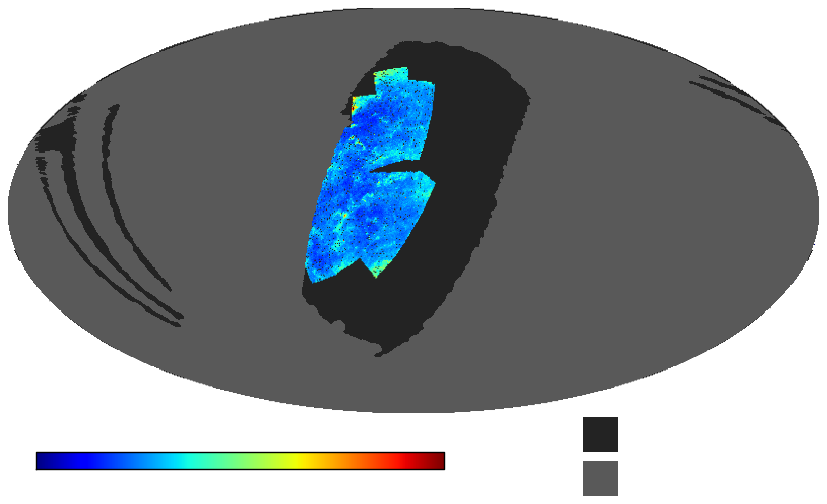}}%
\gplfronttext
\end{picture}%
\endgroup

%% file: cross_spectrum.tex
\begingroup
  \footnotesize
  \selectfont
\makeatletter
\providecommand\color[2][]{%
\GenericError{(gnuplot) \space\space\space\@spaces}{%
Package color not loaded in conjunction with
terminal option `colourtext'%
}{See the gnuplot documentation for explanation.%
}{Either use 'blacktext' in gnuplot or load the package
color.sty in LaTeX.}%
\renewcommand\color[2][]{}%
}%
\providecommand\includegraphics[2][]{%
\GenericError{(gnuplot) \space\space\space\@spaces}{%
Package graphicx or graphics not loaded%
}{See the gnuplot documentation for explanation.%
}{The gnuplot epslatex terminal needs graphicx.sty or graphics.sty.}%
\renewcommand\includegraphics[2][]{}%
}%
\providecommand\rotatebox[2]{#2}%
\@ifundefined{ifGPcolor}{%
\newif\ifGPcolor
\GPcolortrue
}{}%
\@ifundefined{ifGPblacktext}{%
\newif\ifGPblacktext
\GPblacktexttrue
}{}%
\let\gplgaddtomacro\g@addto@macro
\gdef\gplbacktext{}%
\gdef\gplfronttext{}%
\makeatother
\ifGPblacktext
\def\colorrgb#1{}%
\def\colorgray#1{}%
\else
\ifGPcolor
\def\colorrgb#1{\color[rgb]{#1}}%
\def\colorgray#1{\color[gray]{#1}}%
\expandafter\def\csname LTw\endcsname{\color{white}}%
\expandafter\def\csname LTb\endcsname{\color{black}}%
\expandafter\def\csname LTa\endcsname{\color{black}}%
\expandafter\def\csname LT0\endcsname{\color[rgb]{1,0,0}}%
\expandafter\def\csname LT1\endcsname{\color[rgb]{0,1,0}}%
\expandafter\def\csname LT2\endcsname{\color[rgb]{0,0,1}}%
\expandafter\def\csname LT3\endcsname{\color[rgb]{1,0,1}}%
\expandafter\def\csname LT4\endcsname{\color[rgb]{0,1,1}}%
\expandafter\def\csname LT5\endcsname{\color[rgb]{1,1,0}}%
\expandafter\def\csname LT6\endcsname{\color[rgb]{0,0,0}}%
\expandafter\def\csname LT7\endcsname{\color[rgb]{1,0.3,0}}%
\expandafter\def\csname LT8\endcsname{\color[rgb]{0.5,0.5,0.5}}%
\else
\def\colorrgb#1{\color{black}}%
\def\colorgray#1{\color[gray]{#1}}%
\expandafter\def\csname LTw\endcsname{\color{white}}%
\expandafter\def\csname LTb\endcsname{\color{black}}%
\expandafter\def\csname LTa\endcsname{\color{black}}%
\expandafter\def\csname LT0\endcsname{\color{black}}%
\expandafter\def\csname LT1\endcsname{\color{black}}%
\expandafter\def\csname LT2\endcsname{\color{black}}%
\expandafter\def\csname LT3\endcsname{\color{black}}%
\expandafter\def\csname LT4\endcsname{\color{black}}%
\expandafter\def\csname LT5\endcsname{\color{black}}%
\expandafter\def\csname LT6\endcsname{\color{black}}%
\expandafter\def\csname LT7\endcsname{\color{black}}%
\expandafter\def\csname LT8\endcsname{\color{black}}%
\fi
\fi
\setlength{\unitlength}{0.0500bp}%
\begin{picture}(8640.00,4320.00)%
\gplgaddtomacro\gplbacktext{%
\csname LTb\endcsname%
\put(990,1614){\makebox(0,0)[r]{\strut{} 1e-08}}%
\put(990,2675){\makebox(0,0)[r]{\strut{} 1e-07}}%
\put(990,3736){\makebox(0,0)[r]{\strut{} 1e-06}}%
\put(5354,1291){\makebox(0,0){\strut{}}}%
\put(220,2343){\rotatebox{-270}{\makebox(0,0){\strut{}Power ($\mega\jansky\,\steradian^{-1}$)}}}%
}%
\gplgaddtomacro\gplfronttext{%
\csname LTb\endcsname%
\put(7256,3882){\makebox(0,0)[r]{\strut{}CIB + cluster bias model ($\chi^2$/dof = 0.86)}}%
\csname LTb\endcsname%
\put(7256,3662){\makebox(0,0)[r]{\strut{}Power Law ($\chi^2$/dof = 0.63)}}%
}%
\gplgaddtomacro\gplbacktext{%
\csname LTb\endcsname%
\put(990,649){\makebox(0,0)[r]{\strut{}-3e-08}}%
\put(990,954){\makebox(0,0)[r]{\strut{} 0}}%
\put(990,1258){\makebox(0,0)[r]{\strut{} 3e-08}}%
\put(5354,429){\makebox(0,0){\strut{}1000}}%
\put(220,1080){\rotatebox{-270}{\makebox(0,0){\strut{} }}}%
\put(4682,154){\makebox(0,0){\strut{}Multipole ($\ell$)}}%
\put(5751,1296){\makebox(0,0)[l]{\strut{}Null Test ($\chi^2$/dof = 1.4)}}%
}%
\gplgaddtomacro\gplfronttext{%
}%
\gplbacktext
\put(0,0){\includegraphics{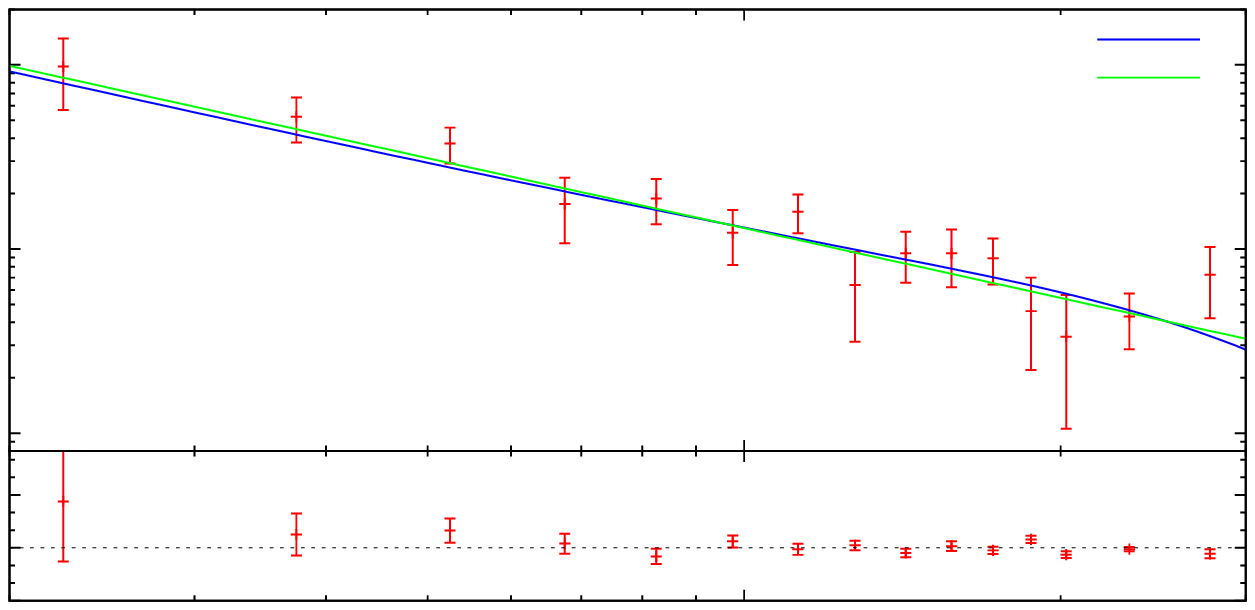}}%
\gplfronttext
\end{picture}%
\endgroup

%% file: vary_masks.tex
\begingroup
  \footnotesize
  \selectfont
\makeatletter
\providecommand\color[2][]{%
\GenericError{(gnuplot) \space\space\space\@spaces}{%
Package color not loaded in conjunction with
terminal option `colourtext'%
}{See the gnuplot documentation for explanation.%
}{Either use 'blacktext' in gnuplot or load the package
color.sty in LaTeX.}%
\renewcommand\color[2][]{}%
}%
\providecommand\includegraphics[2][]{%
\GenericError{(gnuplot) \space\space\space\@spaces}{%
Package graphicx or graphics not loaded%
}{See the gnuplot documentation for explanation.%
}{The gnuplot epslatex terminal needs graphicx.sty or graphics.sty.}%
\renewcommand\includegraphics[2][]{}%
}%
\providecommand\rotatebox[2]{#2}%
\@ifundefined{ifGPcolor}{%
\newif\ifGPcolor
\GPcolortrue
}{}%
\@ifundefined{ifGPblacktext}{%
\newif\ifGPblacktext
\GPblacktexttrue
}{}%
\let\gplgaddtomacro\g@addto@macro
\gdef\gplbacktext{}%
\gdef\gplfronttext{}%
\makeatother
\ifGPblacktext
\def\colorrgb#1{}%
\def\colorgray#1{}%
\else
\ifGPcolor
\def\colorrgb#1{\color[rgb]{#1}}%
\def\colorgray#1{\color[gray]{#1}}%
\expandafter\def\csname LTw\endcsname{\color{white}}%
\expandafter\def\csname LTb\endcsname{\color{black}}%
\expandafter\def\csname LTa\endcsname{\color{black}}%
\expandafter\def\csname LT0\endcsname{\color[rgb]{1,0,0}}%
\expandafter\def\csname LT1\endcsname{\color[rgb]{0,1,0}}%
\expandafter\def\csname LT2\endcsname{\color[rgb]{0,0,1}}%
\expandafter\def\csname LT3\endcsname{\color[rgb]{1,0,1}}%
\expandafter\def\csname LT4\endcsname{\color[rgb]{0,1,1}}%
\expandafter\def\csname LT5\endcsname{\color[rgb]{1,1,0}}%
\expandafter\def\csname LT6\endcsname{\color[rgb]{0,0,0}}%
\expandafter\def\csname LT7\endcsname{\color[rgb]{1,0.3,0}}%
\expandafter\def\csname LT8\endcsname{\color[rgb]{0.5,0.5,0.5}}%
\else
\def\colorrgb#1{\color{black}}%
\def\colorgray#1{\color[gray]{#1}}%
\expandafter\def\csname LTw\endcsname{\color{white}}%
\expandafter\def\csname LTb\endcsname{\color{black}}%
\expandafter\def\csname LTa\endcsname{\color{black}}%
\expandafter\def\csname LT0\endcsname{\color{black}}%
\expandafter\def\csname LT1\endcsname{\color{black}}%
\expandafter\def\csname LT2\endcsname{\color{black}}%
\expandafter\def\csname LT3\endcsname{\color{black}}%
\expandafter\def\csname LT4\endcsname{\color{black}}%
\expandafter\def\csname LT5\endcsname{\color{black}}%
\expandafter\def\csname LT6\endcsname{\color{black}}%
\expandafter\def\csname LT7\endcsname{\color{black}}%
\expandafter\def\csname LT8\endcsname{\color{black}}%
\fi
\fi
\setlength{\unitlength}{0.0500bp}%
\begin{picture}(8640.00,3600.00)%
\gplgaddtomacro\gplbacktext{%
\csname LTb\endcsname%
\put(990,704){\makebox(0,0)[r]{\strut{} 1e-08}}%
\put(990,1695){\makebox(0,0)[r]{\strut{} 1e-07}}%
\put(990,2686){\makebox(0,0)[r]{\strut{} 1e-06}}%
\put(3253,484){\makebox(0,0){\strut{} 1000}}%
\put(352,1931){\rotatebox{-270}{\makebox(0,0){\strut{}Power ($\mega\jansky\,\steradian^{-1}$)}}}%
\put(4652,154){\makebox(0,0){\strut{}Multipole ($\ell$)}}%
\put(2936,3379){\makebox(0,0){\strut{}\emph{Varying IRIS mean intensity mask cut}}}%
}%
\gplgaddtomacro\gplfronttext{%
\csname LTb\endcsname%
\put(4193,2986){\makebox(0,0)[r]{\strut{}Fiducial $2.0\,\mega\jansky\,\steradian^{-1}$ cut}}%
\csname LTb\endcsname%
\put(4193,2766){\makebox(0,0)[r]{\strut{}$1.65\,\mega\jansky\,\steradian^{-1}$ cut}}%
\csname LTb\endcsname%
\put(4193,2546){\makebox(0,0)[r]{\strut{}$2.5\,\mega\jansky\,\steradian^{-1}$ cut}}%
\csname LTb\endcsname%
\put(4193,2326){\makebox(0,0)[r]{\strut{}$3.0\,\mega\jansky\,\steradian^{-1}$ cut}}%
}%
\gplgaddtomacro\gplbacktext{%
\csname LTb\endcsname%
\put(4620,704){\makebox(0,0)[r]{\strut{}}}%
\put(4620,1695){\makebox(0,0)[r]{\strut{}}}%
\put(4620,2686){\makebox(0,0)[r]{\strut{}}}%
\put(6879,484){\makebox(0,0){\strut{} 1000}}%
\put(8279,154){\makebox(0,0){\strut{} }}%
\put(6563,3379){\makebox(0,0){\strut{}\emph{Varying point source mask radius}}}%
}%
\gplgaddtomacro\gplfronttext{%
\csname LTb\endcsname%
\put(7817,2986){\makebox(0,0)[r]{\strut{}Fiducial $5\arcminute$ Radius}}%
\csname LTb\endcsname%
\put(7817,2766){\makebox(0,0)[r]{\strut{}$2\arcminute$ Radius}}%
\csname LTb\endcsname%
\put(7817,2546){\makebox(0,0)[r]{\strut{}$10\arcminute$ Radius}}%
}%
\gplbacktext
\put(0,0){\includegraphics{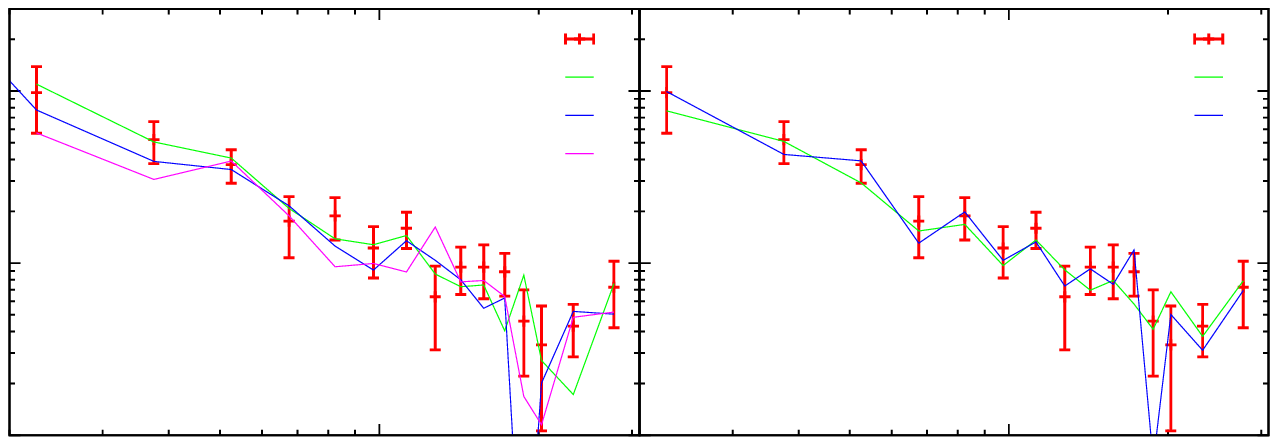}}%
\gplfronttext
\end{picture}%
\endgroup